\documentclass[aps,twocolumn,showpacs,groupedaddress,footinbib]{revtex4-2}

\usepackage{latexsym}
\usepackage{amsmath}
\usepackage{amsfonts}
\usepackage{amssymb}
\usepackage{filecontents} 
\usepackage{graphicx}
\usepackage{svg}
\usepackage{bm}
\usepackage{color}
\usepackage{titlesec}
\usepackage{float}
\usepackage{units}
\usepackage{appendix}

\renewcommand{\vec}[1]{\mathbf{#1}}

\begin{document}
\graphicspath{{plots/}}

\title{Activity-driven phase transition causes coherent flows of chromatin}

\author{Iraj Eshghi}
\author{Alexandra Zidovska} 
\author{Alexander Y. Grosberg}
\email[Corresponding author: ]{ayg1@nyu.edu}
\affiliation{\mbox{Center for Soft Matter Research, Department of Physics, New York University, New York, New York 10003, USA}}

\date{\today}

\begin{abstract}
We discover a new type of nonequilibrium phase transition in a model of chromatin dynamics, which accounts for the coherent motions that have been observed in experiment. The coherent motion is due to the long-range cooperation of molecular motors tethered to chromatin. Cooperation occurs if each motor acts simultaneously on the polymer and the surrounding solvent, exerting on them equal and opposite forces. This drives the flow of solvent past the polymer, which in turn affects the orientation of nearby motors and, if the drive is strong enough, an active polar (``ferromagnetic'') phase of motors can spontaneously form. Depending on boundary conditions, either transverse flows, or sustained longitudinal oscillations and waves are possible. Predicted time and length scales are consistent with experiments. We now have in hand a coarse-grained description of chromatin dynamics which reproduces the  directed coherent flows of chromatin  seen in experiments. This field-theoretic description can be analytically coupled to other features of the nuclear environment such as fluctuating or porous boundaries, local heterogeneities in the distribution of chromatin or its activity, leading to insights on the effects of activity on the cell nucleus and its contents.
\end{abstract}

\maketitle

\textit{Introduction - } Chromatin is the functional form of DNA in living cells, with a variety of active processes such as transcription, replication and DNA repair, taking place directly on the chromatin fiber \cite{alberts2017molecular,van2012chromatin, milo2015cell}. Active forces from these processes affect the organization and dynamics of chromatin \cite{cremer2001chromosome,hubner2010chromatin,zidovska2020self}. Through Displacement Correlation Spectroscopy (DCS), chromatin motions were simultaneously mapped across the entire nucleus in live cells, revealing that chromatin exhibits fast uncorrelated motions at short times ($<$ 1 s) and slow correlated motions at longer times \cite{zidovska2013micron}. The correlated chromatin motions are coherent over 3--5 $\mu$m for several seconds, before the coherent domains break up and new ones form, resembling an oscillatory-like behavior \cite{zidovska2013micron}. Furthermore, while the uncorrelated motions were shown to be thermal-like, the coherent chromatin flows were eliminated upon ATP depletion or inhibition of major nuclear enzymes such as RNA polymerase II, DNA polymerase and topoisomerase II, demonstrating active, energy-dissipating and nonequilibrium nature of the coherent chromatin flows \cite{zidovska2013micron,bruinsma2014chromatin,saintillan2018extensile}. \\
\indent From the active matter prospective, hydrodynamics of systems with activity was the subject of many studies, as reviewed in \cite{marchetti2013hydrodynamics}. Depending on the role of solvent and the symmetry of the order parameter \cite{vicsek1995novel,toner1995long}, active hydrodynamics exhibit phenomena ranging from coherent instabilities \cite{simha2002hydrodynamic,kruse2004asters}, to nematic or polar order \cite{ahmadi2005nematic,AdarJoanny2021}, to treadmilling \cite{julicher2007active, callan2011hydrodynamics}.  In many works, e.g., on active nematics the idea is that nematic order is formed as in the usual passive system, due to interactions between, say, elongated molecules, and then activity drives spectacularly interesting dynamics (see \cite{keber2014topology}).\\
\indent In the context of chromatin, molecular motors driving active dynamics, such as RNA polymerases, do not appear to be close enough to form a long-range order due to direct contact with each other \cite{zhao2014spatial}. At the same time, hydrodynamic treatment of chromatin finds that coherent chromatin dynamics can be sustained only in the presence of the ordered orientations of force dipoles \cite{bruinsma2014chromatin}. In alternative hydrodynamics-free approaches, computationally reproducing coherent chromatin motions required the use of artificial long-range interactions \cite{shi2018interphase,liu2018chain,di2018anomalous}. An important hint came from hydrodynamic simulations work, where large-scale coherent chromatin dynamics as well as strong nematic order of chromatin fiber was observed, without inserting any artificial long-range forces \cite{saintillan2018extensile}. Instead, this model relies on the non-specific effects of hydrodynamics to mediate such interactions. In our earlier study, we identified motors, which exert equal but opposite forces on the polymer and solvent, as responsible for the large-scale hydrodynamic flows in the chromatin-nucleoplasm two-fluid system \cite{eshghi2022symmetry}. Here, we aim to develop a coarse-grained hydrodynamic model, which reproduces the development of the coherent chromatin phase. We hypothesize that there can be an ordering phase transition when the force of the motors exceeds a threshold value. We seek to analyze which properties of the chromatin-nucleoplasm system govern this phase transition as well as the structure of ordered phase.  \\

\begin{figure}
  \centering
  \includegraphics[width=\linewidth]{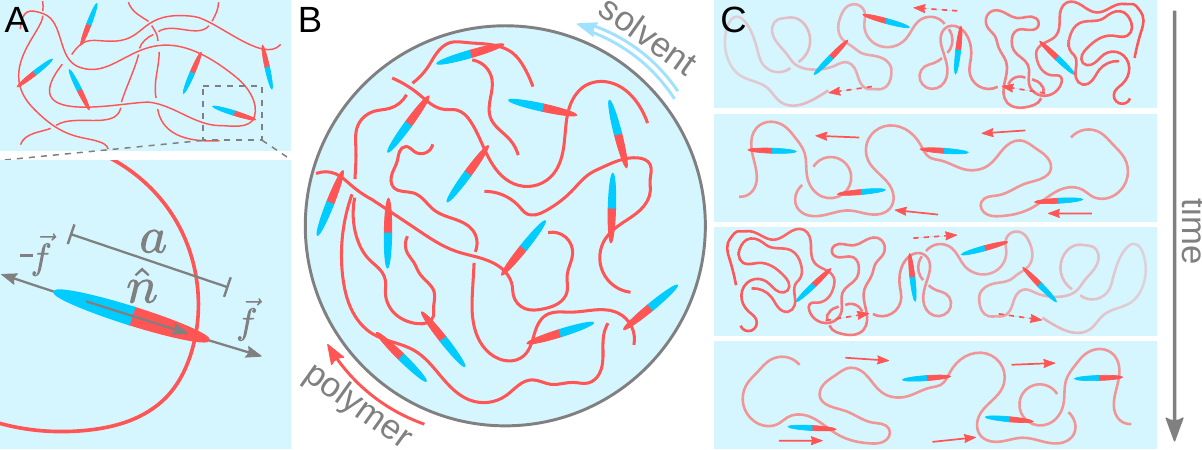}\\
  \caption{Sketch of our model and the two types of solutions. A: Example of a region of disordered polymer and attached force dipoles, with a zoomed-in section where the parameters describing the microscopic features of the motors are shown. B: Sketch of the transverse solution in a spherical domain, showing the polar alignment of the sources and the sustained solvent flow being pumped in the opposite direction of their orientation. C: Sketch of the longitudinal, oscillatory solution to the equations of motion. Dashed arrows show the relaxational (osmotic) flow of polymer in the absence of active forces, which seeks to even out density fluctuations. Solid arrows show the active polymer flow induced by the sources. Time goes from the upper panel to the lower one, with time per frame given in Eq. (\ref{eq:geometric_mean}).}
  \label{fig:model_and_solutions}
\end{figure}

\textit{The model and equations of motion: linear response - }  Following earlier work \cite{bruinsma2014chromatin,eshghi2022symmetry} we describe chromatin using the two-fluid model originally by Doi and Onuki \cite{doi1992dynamic}. The dynamics of the system in this model is described by the fields of polymer velocity $\vec v^{\mathrm{p}}(\vec r,t)$, polymer volume fraction $\phi(\vec r,t)$, and the solvent velocity $\vec v^{\mathrm{s}}(\vec r,t)$, while the solvent volume fraction is $1-\phi(\vec r,t)$ because of overall incompressibility. To describe the onset of spontaneous symmetry breaking and formation of polar  ordered domains, we start with the assumption of linear and local rheological response of the polymer.  This implies that the velocities are small, as are the deviations from the average density, $\phi(\vec r,t) = \phi_0 + \delta\phi(\vec r,t)$.  This implies further that polymer osmotic pressure is $\Pi \simeq K \delta \phi(\vec r,t)$, with osmotic modulus $K$, while  the force resulting from  polymer viscous stress is $\eta^{\mathrm{p}} \star \nabla^2 \vec v^{\mathrm{p}} (\vec r, t)$, where polymer viscosity may have some time memory kernel and $\star$ means convolution  (see below about neglect of extensional viscosity and terms $\sim \nabla(\nabla\cdot \vec v^{\mathrm{p}})$). In this approximation, equations of motion of the model are conveniently written in the Fourier-transformed frequency domain (with sign convention $\partial /\partial t \to - i \omega$) as follows:
\begin{subequations}
\label{eq:eqmot_fluid1}
\begin{align}
\zeta(\vec v^{\mathrm{p}}_{\omega} - \vec v^{\mathrm{s}}_{\omega}) &= \eta^{\mathrm{p}}_{\omega}\nabla^2\vec v^{\mathrm{p}}_{\omega} -K \nabla \delta\phi_{\omega} - \phi_0\nabla P_{\omega} + \vec F^{\mathrm{p}}_{\omega}\\
\zeta(\vec v^{\mathrm{s}}_{\omega} - \vec v^{\mathrm{p}}_{\omega}) &= \eta^{\mathrm{s}}\nabla^2\vec v^{\mathrm{s}}_{\omega} - (1-\phi_0)\nabla P_{\omega} + \vec F^{\mathrm{s}}_{\omega}\\
i\omega \delta\phi_{\omega} &= \phi_0\nabla\cdot \vec v^{\mathrm{p}}_{\omega} = - (1-\phi_0) \nabla\cdot \vec v^{\mathrm{s}}_{\omega}
\end{align}
\end{subequations}
The first two equations represent force balance conditions for polymer and solvent respectively, while the last two are continuity conditions for these two components. Here $\zeta$ is the friction coefficient of polymer against solvent, per unit volume, $\eta^{\mathrm{s}}$ is the viscosity of the solvent, $P_{\omega}$ is the hydrostatic pressure.\\
\indent 
The heart of the problem is the understanding of active force densities $\vec F^{\mathrm{p}}$ and $\vec F^{\mathrm{s}}$ generated by active motors. Typical size of every motor, which we denote $a$, is on the order of or smaller than the mesh size $\lambda$. 
As explained above, we focus on motors exerting equal and opposite forces on polymer and on solvent, which to the first approximation, means $\vec F^{\mathrm{p}} = -\vec F^{\mathrm{s}} = f\rho \vec m(\vec r,t)$, 
%
%
where $\rho$ is the number density of motors,  while $\vec m(\vec r,t) = \left< \hat{\vec n} \right>$ is the average orientation. With $f>0$, this describes extensile force dipoles, contractile ones correspond to $f < 0 $. Remaining within linear response, we assume $\left| \vec m \right|$ small and neglect the change of motor density associated with changing polymer density $\delta \phi$. Note that every motor has, generally, some finite processivity, stemming from its on- and off-rates; density $\rho$ includes only those motors that are simultaneously working. The geometry of the source is illustrated in Fig. \ref{fig:model_and_solutions}A.

Since the body of the force-exerting motor is tethered to the polymer at one end and experiences friction from the solvent, there must be a torque acting on the motor and proportional to the relative velocity $\vec v^{\mathrm{p}} - \vec v^{\mathrm{s}} = \vec w$, leading to the following dynamics of the $\vec m$ field (see Appendix, Section \ref{sec:singlemotor} for detailed derivation):
\begin{equation}
    - i \omega \vec m_{\omega}  = \frac{2}{3a}\vec w_{\omega} - 2\frac{T}{\gamma}\vec m_{\omega} \ ,
    \label{eq:eqmot_mag_linear}
\end{equation}
where $\gamma$ is the rotational drag coefficient for the motor.\\
\indent    
 Apart from nonlinearities (considered below), we neglect in Eq. (\ref{eq:eqmot_mag_linear}) coupling of motor orientation to polymer concentration gradient (because motor size is smaller than or comparable to polymer mesh); don't consider renormalization of active force due to the flow itself (because $f$ is large enough); ignore the possibility of the induced nematicity of the polymer and corresponding active stress.  Our theory is in some ways similar to that of Adar and Joanny \cite{AdarJoanny2021}, as they also examine coupling between flow and polarization in a two fluid model, but they focus on the regime of strong polarization which can only rotate in response to the flow, while we concentrate on the chromatin-relevant opposite regime of weak polarization which only arises due to the flow. \\
\indent 
Along with Eq. (\ref{eq:eqmot_mag_linear}), it is convenient to recast the equations of motion (\ref{eq:eqmot_fluid1}) in terms of the above defined relative velocity $\vec w_{\omega} = \vec v^{\mathrm{p}}_{\omega}-\vec v^{\mathrm{s}}_{\omega}$ and viscosity-weighted average velocity $\vec u_{\omega} = \left. \left(\eta^{\mathrm{p}}_{\omega}\vec v^{\mathrm{p}}_{\omega}+\eta^{\mathrm{s}}\vec v^{\mathrm{s}}_{\omega}\right)  \right/ \left(\eta^{\mathrm{p}}_{\omega}+\eta^{\mathrm{s}}\right)$  (see Appendix, Section \ref{sec:full_derivation}). 
Doing so, one can easily see that  relative velocity $\mathbf{w}$ is driven by $\mathbf{m}$, i.e., mathematically by force monopoles rather than dipoles.  A similar mathematical structure appeared in the work \cite{kumar2014flocking}, albeit in an entirely different physics context.  This explains why hydrodynamic interactions are so important in our active system, despite the fact that in passive polymers they are screened at the distances not far exceeding the mesh size \cite{Everaers_HD_Interaction}.  Another remarkable feature of the full set of equations is that they 
allow for simultaneous Helmholtz decomposition of the three vector fields $\vec m$, $\vec u$, and $\vec w$ to the uncoupled divergence-free (transverse, $\perp$) and curl-free (longitudinal, $\parallel$) modes.

\textit{Threshold of instability, divergence free (transverse) modes - }  Transverse modes do not involve density change, $\delta \phi = 0$, and, accordingly, no pressure gradient, $\nabla P = 0$.  This leaves us with just two equations which are easily combined into one (see Appendix, Section \ref{sec:full_derivation}):\small
\begin{equation}
    -i\omega\tau\left(1 - \lambda^2\nabla^2\right)\vec w_{\omega\perp} = 2\left(\frac{f\rho\gamma}{3a\zeta T} - 1 + \lambda^2\nabla^2\right)\vec w_{\omega\perp}\; ,
    \label{eq:transverse_flows}
\end{equation}
\normalsize where we introduced short hand notations
\begin{equation} \tau = \frac{\gamma}{T} \ , \ \ \text{and} \ \ \ \lambda^{2} = \frac{\eta_{\omega}^{\mathrm{p}} \eta^{\mathrm{s}} /\zeta}{\eta_{\omega}^{\mathrm{p}} + \eta^{\mathrm{s}}} \simeq \frac{\eta^{\mathrm{s}}}{\zeta} \ , \label{eq:definition_tau_lambda} \end{equation}
and in the last transformation we took into account the fact that $\eta_{\omega}^{\mathrm{p}} \gg \eta^{\mathrm{s}}$, by several orders of magnitude, over the entire frequency range of interest \cite{tseng2004micro,hameed2012dynamics,celedon2011magnetic,de2007direct,eshghi2021interphase,caragine2018surface,liang2009noninvasive,erdel2015viscoelastic,zidovska2020rich}.  Clearly, $\tau$ is the characteristic time of passive re-orientation by a single motor, while $\lambda$ is the length scale of the mesh size.

In an infinite domain, the modes are just plane waves, $\nabla^2 \to - q^2$, and we see that modes become unstable when $\frac{f \rho \gamma}{3 a \zeta T} > 1+ \lambda^2 q^2$.  The fact that the length scale $1/q$ of the unstable modes diverges as we approach from above the critical force level at which $\frac{f\rho\gamma}{3a\zeta T}-1 = 0$ is reminiscent of a second-order phase transition, similar to that in a magnet, with $\vec w_{\perp}$ playing the role of (self-consistent) magnetic field and $\vec m_{\perp}$ the local averaged spin.  The critical parameter $\epsilon = \frac{f\rho\gamma}{3a\zeta T}-1 $ describes a competition between the velocity produced by the cooperatively acting motors $ \frac{f\rho}{\zeta}$, and the characteristic velocity needed to align a motor, $a T/\gamma = a/\tau$.

\begin{figure}[b]
    \centering
    \includegraphics[width=\linewidth]{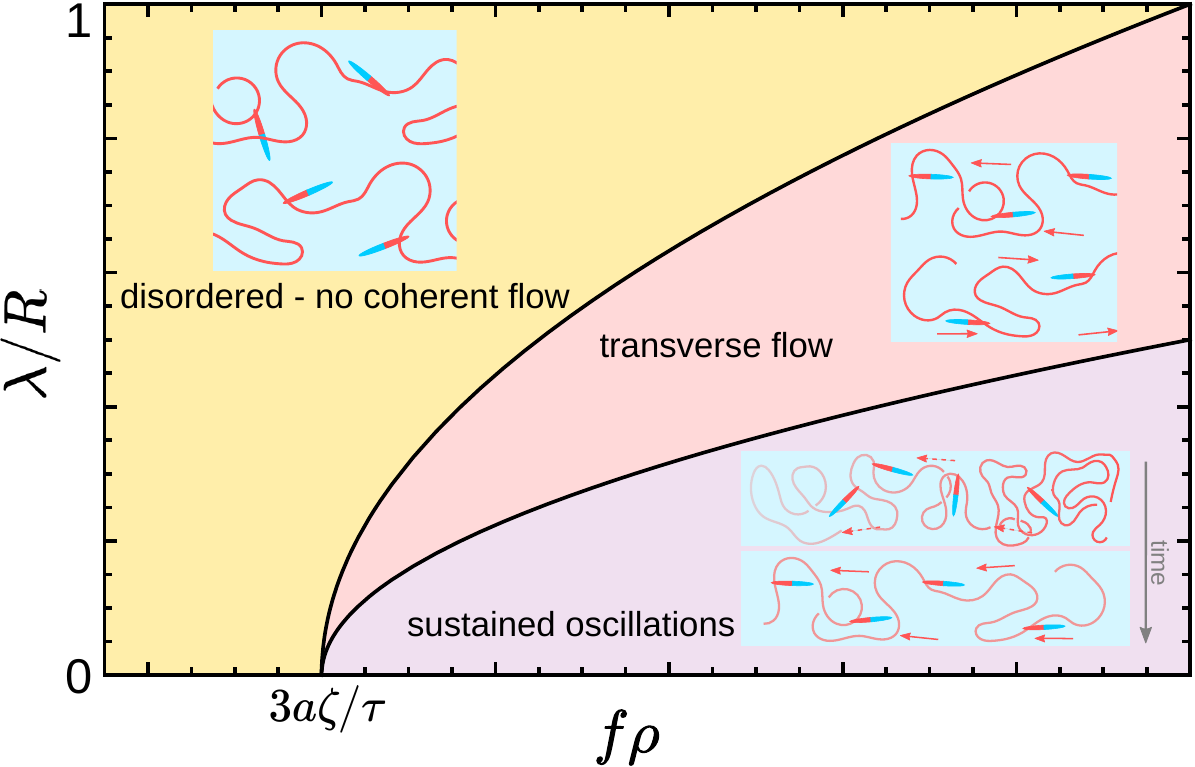}
    \caption{Phase diagram of the instabilities and the regions of parameter space where they develop, as a function of the active force density and the size of the container. In the left (blue) region of the diagram, the forcing is insufficient to drive instabilities and the system remains disordered. The middle (red) region is where the forces are sufficient to drive transverse flows but not strong enough to cause polymer density fluctuations. Finally, the bottom-right (purple) region of parameter space is where both longitudinal oscillations and transverse flows are possible. The lines separating the regions correspond to the conditions (\ref{eq:threshold_transverse},\ref{eq:threshold_longitudinal}) respectively.}
    \label{fig:phase_diagram}
\end{figure}

If the system is confined in a finite domain of size $R$, then modes have a more elaborate structure and discrete spectrum.   Although the stability analysis for this case may require a separate study \cite{eshghi2023model}, the qualitative estimate of the amount of force needed to generate instability can be obtained by just setting $q \sim 1/R$ (see Fig. \ref{fig:phase_diagram}):
\begin{equation} f \rho > \frac{3 a \zeta T}{\gamma} + \frac{a \eta_s T}{\gamma R^2} \ .  \label{eq:threshold_transverse}\end{equation}
The meaning of this condition becomes transparent if we imagine an arrangement of motors in a typical transverse mode in a round domain (depicted in Fig. \ref{fig:model_and_solutions}B).  These motors acting together have to be strong enough to overcome the friction of the solvent pumped through the network (the first term) and additional friction against the boundary (the second term).

\textit{Threshold of instability, curl free (longitudinal) modes - }  The longitudinal waves involve density fluctuations, which is why their description is more complicated.  Nevertheless, even in this case, the problem is reduced to a single equation for the field $\vec w_{\parallel}$ (see Appendix, Section \ref{sec:full_derivation} for derivation):
\begin{equation}
\begin{split}
    & \left[ 1-\lambda_s^2\nabla^2 \right] \tau^2\partial_t^2\vec w_{\parallel} -  4 \lambda_d^2 \nabla^2\vec w_{\parallel} + \\ & + 2 \left[ 1  - \left( \lambda_s^2+\lambda_d^2 \right)\nabla^2 - \frac{f \rho \gamma}{3 a \zeta T} \right] \tau\partial_t\vec w_{\parallel} = 0 \ ,
    \label{eq:eqmot_longitudinal}
\end{split}
\end{equation}
where in addition to (\ref{eq:definition_tau_lambda}) we introduced two new length scales, their complete expressions are cumbersome (see Appendix, Eq. \ref{eq:lengthscales}), but in simplified form  (due to $\eta^{\mathrm{p}}_{\omega} \gg \eta^{\mathrm{s}}$) they are as follows:
\begin{equation} \lambda_s^2 \simeq \frac{  \eta^{\mathrm{p}} \left(1- \phi_0 \right)^2 }{\zeta }  \ \ \text{and} \ \ \lambda_d^2 \simeq \frac{K \phi_0 \left( 1 - \phi_0 \right)^2    \gamma}{2 \zeta T} \ .
\end{equation}
In equation (\ref{eq:eqmot_longitudinal}), we returned to time domain ($-i \omega \to \partial_{t}$), making the oscillator structure of the equation more transparent. This is possible only as long as polymer viscosity, $\eta^{\mathrm{p}}_{\omega}$, is only smoothly dependent on frequency.  

As in the transverse case before, in an infinite domain the modes are just plane waves, $\nabla^2 \to - q^2$, and Eq. (\ref{eq:eqmot_longitudinal}) becomes that of a damped harmonic oscillator.  Remarkably, active driving force comes only in the friction term.  In particular, sufficiently strong and numerous motors can lead to the flipped sign of friction, making the oscillator unstable.  As before,  structure of modes for a finite size domain of size $R$ requires special analysis \cite{eshghi2023model}, but qualitatively we can estimate the instability threshold by just replacing $q \to 1/R$ (see Fig. \ref{fig:phase_diagram}):
\begin{equation}   f \rho  > \frac{3 a \zeta T}{\gamma}  + \frac{(1-\phi_0)^2 a}{R^2} \left[ \frac{3 T \eta^{\mathrm{p}}}{\gamma} + \frac{3}{2} K \phi_0 \right]  \ .  \label{eq:threshold_longitudinal} \end{equation}
Similar to the formula (\ref{eq:threshold_transverse}) for the transverse case, Eq. (\ref{eq:threshold_longitudinal}) means that motors have to be strong enough to overcome friction, which this time involves moving and deforming polymers, thus dependent on $\eta_{\omega}^{\mathrm{p}}$ and $K$, respectively.  This implies that a larger force is needed to generate longitudinal modes compared to the transverse ones  (and the extensional viscosity of the polymer can further increase this threshold). 

When force is exactly equal to the threshold value for some $q$, this mode exhibits a sustained oscillation with frequency such that $\left( \omega \tau \right)^2 = 2 \lambda_d^2 q^2 / \left( 1 + \lambda_s ^2 q^2 \right)$.  In particular, the small $q$ modes ($q \lambda_s \ll 1$) are just propagating waves with $\omega \propto q$ and with velocity $\sim \lambda_d/\tau \sim K/\zeta$.  Numerically generated movies illustrating possible wave packet dynamics can be found in Appendix, Section \ref{sec:numerics}.

This can be rationalized in an interesting way.  Let us define rate $\tau_{q}^{-1} \sim (K/\zeta ) q^2$; given that $K/\zeta$ has dimensionality of a diffusion coefficient, $\tau_q$ is the characteristic relaxation time of a density wave of length $1/q$ by cooperative diffusion, driven by polymer elasticity ($K$) against friction ($\zeta$).  In terms of $\tau_q$, we can write mode $q$ frequency as the geometric mean of two rates:
\begin{equation} \omega \sim \left(\tau \tau_{q} \right)^{-1/2} \ , \ \ \mathrm{with} \ \ \tau_{q}^{-1} \sim (K/\zeta ) q^2 \ . \label{eq:geometric_mean} \end{equation}
The mathematical structure of frequency as the geometric mean of two rates is analogous to that which arises in the Lotka-Volterra equations 
\cite{lotka1910contribution,volterra1928variations}, which is the geometric mean of the growth rate of the prey and the death rate of the predator. This structure reflects the physical nature of the oscillator: by the time some dense region of size $1/q$ relaxes, it will have generated a velocity field which locally aligns the field $\vec m_{\parallel}$. This field  has a persistence time $\tau$, and pumps the polymer in the same direction in which it was relaxing. This causes a new dense region to develop, until the dipoles lose their alignment in turn after a time $\tau$, and the polymer relaxation begins yet again at a rate $1/\tau_q$ in the opposite direction. This is illustrated in Fig. \ref{fig:model_and_solutions}C. 

If the force is slightly above or slightly below the threshold (\ref{eq:threshold_longitudinal}), then oscillator is either slowly decaying (below) or slowly increase swinging (above), with characteristic time that diverges at the threshold, again reminiscent of a standard critical slowing down in phase transitions.

\textit{Beyond linear response - }  Once driving force exceeds the threshold value, unstable modes exponentially explode, grow out of the linear response range, and then non-linearity comes to rescue and eventually arrests the growth.  There are many non-linear effects possible, including non-linear osmotic and/or rheological behavior, advection of motors, but we will focus on the most basic and omnipresent one, namely, the fact that orientational order of motors is limited such that $\left| \vec m \right| \leq 1$: the maximum motors can do together is to align completely.\\
\indent 
Complete description of orientation dynamics in an orienting field is rather cumbersome (see Appendix, Section \ref{sec:singlemotor}).  We will restrict ourselves with the simplest estimate, assuming that polarization vector $\vec m$ beyond linear regime (\ref{eq:eqmot_mag_linear}) evolves according to
\begin{equation}
\begin{split}
\tau \partial_t \vec m   & = 2 \left( \vec m_{\mathrm{eq}} \left( \vec w \right) - \vec m \right) \ , \\ & \ \ \ \ \mathrm{with} \ \vec m_{\mathrm{eq}} \left( \vec w \right)  \simeq \vec w \frac{\tau}{3a} \left( 1 - \frac{\left( \vec w \tau/a \right)^2}{15} \right) \ .
\end{split}   \label{eq:eqmot_mag_nonlinear}
\end{equation}
Here $\vec m_{\mathrm{eq}} \left( \vec w \right)$ is the equilibrium value that would be achieved in a constant flow $\vec w$; similar to classical orientation of dipoles, $m_{\mathrm{eq}}(w) = \coth\left( w \gamma / aT \right) - a T/\gamma w$, and we use the first non-linear term of expansion.  Eq. (\ref{eq:eqmot_mag_nonlinear}) is not exact, but captures main qualitative features.\\
\indent 
Once the dynamics is nonlinear, separation of longitudinal and transverse modes is not possible.  
 Nevertheless, neglecting frequency dependence of $\eta^{\mathrm{p}}$ (and, therefore, $\lambda_s$), we can reduce equations of motion to a single equation (see Appendix, Section \ref{sec:full_derivation}): 
\begin{equation}
\begin{split}
 & \tau^2 \partial_t^2\left[ 1 - \lambda_s^2 \nabla \nabla \cdot + \lambda^2 \nabla \times \nabla \times \right] \vec w - 4 \left[ \lambda_d^2 \nabla \nabla \cdot \right] \vec w   \\ & + 2 \tau\partial_t \left[ 1  - \left( \lambda_s^2+\lambda_d^2 \right)\nabla \nabla \cdot + \lambda^2 \nabla \times \nabla \times  - \right. \\ & \left. \ \ \ \ \ \ \ \ \ \ \ \ \ \ \ \ \ \ \ \ \ - \frac{ f \rho \gamma}{3 a \zeta T} \left( 1 - \frac{ \tau^2}{15 a^2} \vec w^2 \right) \right] \vec w = 0 \ .
\end{split}
\label{eq:eqmot_nonlinear}
\end{equation}
Equation (\ref{eq:eqmot_nonlinear}) is instructive. First of all, if we drop the nonlinear term, then it is reduced to either Eq. (\ref{eq:transverse_flows}) or Eq. (\ref{eq:eqmot_longitudinal}) if the field $\vec w$ is divergence-free or curl-free, respectively \footnote{It is useful to note parenthetically the following simple fact about signs: in spatial Fourier representation, while $\nabla \nabla \cdot \mathbf{w}  \to - q^2 \mathbf{w}$, with minus sign, but $\nabla \times \nabla \times \mathbf{w}  \to + q^2 \mathbf{w}$, with the sign plus}.  Of course, full non-linear equation is difficult to analyze.  Nevertheless, Eq. (\ref{eq:eqmot_nonlinear}) is still similar to that for an oscillator (specifically, van der Pol oscillator \cite{vanderpol1926,AndronovVittKhaikin}),   with both active forces and non-linear saturation contributing to the friction term (with first time derivative); All types of second spatial derivatives, arising from viscous stresses, are controlled by the domain size and estimated as $1/R^2$, although the detailed shape of the vector field $\vec w$ is sensitive to the domain shape and boundary conditions.  For an estimate, we just say that modes start to grow when force makes friction term in Eq. (\ref{eq:eqmot_nonlinear}) negative and then $\left| \vec w \right|$ grows until friction term becomes positive again.  If the force threshold for instability is $f^{\ast}$ (determined, e.g., by Eq. (\ref{eq:threshold_longitudinal})), then the steady velocity amplitude scales as $w^2 \sim \left( 15 a^2 / \tau^2 \right) \left( f - f^{\ast} \right)/f^{\ast}$, and corresponding density variations amplitude is $\delta \phi^2 \sim \left(15 a^2 \zeta / K \tau \right) \left( f - f^{\ast} \right)/f^{\ast}$. Corresponding numerical solutions are shown in Appendix, Section \ref{sec:numerics}.\\
\indent
\textit{Discussion - } Our model predicts three phases for chromatin dynamics: disordered, and two types of polar order - transverse flows and oscillatory regime. These are controlled by the active force density $f\rho$ and the domain size $R$ (Fig. \ref{fig:phase_diagram}).\\
\indent
Our results are consistent with extensive simulations reported in \cite{saintillan2018extensile}, showing that extensile motors ($f>0$), if present in sufficient density $f\rho$, produce polar ordered state and coherent motion.  An additional feature of the computational model  \cite{saintillan2018extensile} is that they observe nematic ordering of polymer itself; we speculate that nematicity of the polymer may be a consequence of the polar order of motors, because the motors in the simulations were tied to local direction of the polymer.\\
\indent
Speaking about chromatin \textit{in vivo}, we consider RNA polymerase II as a likely motor driving chromatin dynamics, as it binds to chromatin and pushes RNA into the solvent \cite{alberts2017molecular}, although many other nuclear enzymes can also mechanically couple chromatin fiber to the nucleoplasm, e.g., loop extruding condensin \cite{dekker2018condensin}. For these motors, density $\rho \gtrsim \unit[10^2]{\mu m^{-3}}$ \cite{kimura1999quantitation}, force $f \sim \unit[25]{pN}$ \cite{wang1998force}, size $a \sim \unit[20]{nm}$ \cite{rhodin2003single}. At full cooperation, when perfectly aligned, these motors can drive solvent past chromatin at a very large speed $w_{\mathrm{max}} \sim f \rho /\zeta \sim \unit[10^7]{nm / s}$; here, we used $\zeta =  \eta^{\mathrm{s}}/\lambda^2$, assuming nucleoplasm viscosity similar to that of water, $\eta^{\mathrm{s}} \sim \unit[10^{-3}]{Pa \cdot s}$ \cite{liang2009noninvasive,erdel2015viscoelastic}, and taking chromatin mesh size $\lambda \sim \unit[50]{nm}$ ($\unit[30- 100]{nm}$ reported in experiments \cite{gorisch2003diffusion,solovei2002spatial}).  Of course, polymer moves with a smaller speed, reduced by a factor of ratio of viscosities, $v^{\mathrm{p}} \sim \left(\eta^{\mathrm{s}} / \eta^{\mathrm{p}} \right) w$.\\
\indent
Unfortunately, the ratio of viscosities is difficult to measure directly. Using experimentally measured values of $\eta^{\mathrm{p}}$ and $\eta^{\mathrm{s}}$ \cite{tseng2004micro,de2007direct,liang2009noninvasive,celedon2011magnetic,hameed2012dynamics,erdel2015viscoelastic,caragine2018surface,eshghi2021interphase,zidovska2020rich}, we estimate the ratio to be in the range $10^{-2}$ to $10^{-6}$. The latter figure would be in agreement with experimentally measured polymer speed in slow coherent motion about $\unit[10]{nm / s}$ \cite{zidovska2013micron}.  If the actual ratio of viscosities is not quite that small, then we will have to conclude that chromatin \textit{in vivo} operates close to criticality, where our model predicts reduction of velocity by a factor $\left( f\rho / \left(f\rho\right)^{\ast} - 1 \right)^{1/2}$. \\
\indent
To estimate actual closeness to criticality in the case of transverse flows, it is convenient to rewrite the critical conditions Eq.  (\ref{eq:threshold_transverse}) in terms of the above mentioned maximal speed at full cooperation: $f\rho/\zeta  > \left(f\rho\right)^{\ast} /\zeta = (3a/\tau) \left[1 + \lambda^2/R^2 \right] $.  Here $\lambda/R$ is completely negligible for realistic nucleus size of about $R \sim \unit[10]{\mu m}$ \cite{alberts2017molecular,milo2015cell}, while passive reorientation time of a motor we calculate as $\tau \gtrsim \unit[10^{-6}]{s}$ (see Appendix, Section \ref{sec:estimates}). The actual value of $\tau$ could be significantly higher, since we underestimated the dissipative coupling between motor and polymer. Current estimate yields $3a/\tau \lesssim \unit[10^7]{nm / s}$, similar to $w_{\mathrm{max}}$ above.  This suggests that transverse flows could indeed be responsible for the coherent chromatin flows in live cells. \\
\indent
In the oscillatory regime, required critical force density is larger, $f\rho/\zeta  > \left(f\rho\right)^{\ast} /\zeta = (3a/\tau) \left[1 + (\eta^{\mathrm{p}} / \eta^{\mathrm{s}})\lambda^2/R^2 \right] $, see Eq. (\ref{eq:threshold_longitudinal}).  Given the uncertainties in the estimates of $\tau$ and, most importantly, ratio of viscosities, it is difficult at the present time to make definitive statements about feasibility of this regime for \textit{in vivo} chromatin. A similar uncertainty exists about our predictions of running waves speed and oscillations period (Eq. \ref{eq:geometric_mean}), which is poorly constrained, but seems significantly shorter than measured lifetime of coherent chromatin flows in cells of $\sim \unit[5-10]{s}$ \cite{zidovska2013micron}. Importantly, a set of parameters consistent with current knowledge can be chosen that yields physiologically relevant results, yet such a choice cannot be presently motivated.\\
\indent
Overall, our model might be consistent with current measurements, although the significant approximations in our theory and uncertainties in parameters call for future efforts towards more detailed modeling.
 This will require consideration of the boundary conditions \cite{eshghi2023model}, including solvent permeation through the nuclear envelope \cite{paine1975nuclear}, coupling of chromatin to lamin \cite{macpherson2020chromatin,falk2019heterochromatin,Mahajan2022hetero} and to nuclear envelope fluctuations \cite{chu2017origin}. Another promising direction is to account for a nonuniform distribution of active motors in the nucleus and along the chromatin fiber, such as active motors preferentially residing in transcriptionally active euchromatin \cite{shi2018interphase,Mahajan2022hetero,goychuk2022polymer}. But already now our theory makes predictions that beg for experimental tests, in particular for solvent motions, which unlike chromatin motions have not been measured before.\\
\indent 
AZ is grateful for support from the NSF Grants CAREER PHY-1554880, CMMI-1762506, PHY-2210541, DMS-2153432 and NYU MRSEC DMR-1420073. This research was supported in part by the National Science Foundation under Grant No. NSF PHY-1748958. AZ and AYG acknowledge useful discussions with participants of the 2020 virtual KITP program on "Biological Physics of Chromosomes". AYG thanks S. Ramaswamy for useful discussions.
\newpage
\begin{center}
    \textbf{Appendices}
\end{center}

\appendix
\section{Single motor dynamics}
\label{sec:singlemotor}
Since the body of the force-exerting motor is tethered to the polymer at one end and experiences friction from the solvent, there must be a torque acting on the motor and proportional to the relative speed of polymer past solvent, $\vec v^{\mathrm{p}} - \vec v^{\mathrm{s}} = \vec w$. This leads to the following Langevin equation describing the stochastic dynamics of the motor orientation vector $\hat{\vec n}$:
\begin{equation}
\begin{split}
    &\partial_t \hat{\vec n} = \left(\mathcal{I}  - \hat{\vec n}\hat{\vec n}\right)\cdot\left[\frac{\vec w}{a} + \sqrt{\frac{2 T}{\gamma}}\ \pmb{\xi}\right]\\
    &\langle \xi_i(t) \xi_j(t')\rangle = \delta_{ij}\delta(t-t')\;.
    \label{eq:orientation_eqmot}
\end{split}
\end{equation}
We assume here that the dipoles experience rotational friction with coefficient $\gamma$. We also assume the presence of Gaussian white noise that obeys fluctuation-dissipation theorem and thus has variance $2 \gamma T$, with $T$ temperature (although we do not \textit{a priori} exclude the possibility that $T$ may be some sort of an effective temperature). $\mathcal{I}-\hat{\vec n}\hat{\vec n}$ (with $\mathcal{I}$ the identity matrix) projects the expression in square brackets onto the plane perpendicular to $\hat{\vec n}$ and thus ensures that the dynamics does not change the length and only rotates the $\hat{\vec n}$ vector. It is worth noting that equation (\ref{eq:orientation_eqmot}) is identical to the equation of motion for Langevin dipoles in an external electric field at finite temperature, where in our case $\vec w$ plays the role of orienting field \cite{debye1929polar}.

Equation (\ref{eq:orientation_eqmot}) is an embodiment of our minimal model of force dipole dynamics. It certainly neglects a number of potentially relevant factors, two of which we mention.  First, we do not take into account any direct interaction between motors (e.g., excluded volume), assuming they are sufficiently far apart. Second, we assume that motor is attached to a polymer by a swivel and thus turning the motor does not cause polymer to bend; in other words, motor direction $\hat{\vec n}$ is assumed independent of the local direction of the polymer backbone (note that the computational model in the work \cite{saintillan2018extensile} makes essentially the opposite assumption that these two vectors are the same).

To describe the onset of polar order, we consider the dynamics of the coarse-grained orientation field $\vec m(\vec r,t)$.  We define the coarse-graining to take place inside a ball $\mathcal{B}$ centered at $\vec r$, such that this ball is large enough to contain many dipoles, while still being smaller than the relevant dynamic length scales over which gradients develop in the system. Within this ball, we define $\vec m(\vec r) = \frac{\sum \hat{\vec n}_i}{\sum i}$, the average orientation. Thus, $\vec m$ is a vector with length $0\leq |\vec m|\leq 1$.\\
\indent Then, to derive the equation of motion for $\vec m$, we consider the distribution of directions of motors in $\mathcal{B}$. We can assume that the external field $\vec w$ is constant in this region, and we orient our coordinate system such that it points in the $\hat{\vec z}$ direction. Then, the external field leads to an effective potential which in spherical coordinates is proportional to $\cos(\theta)$. The resulting Fokker-Planck equation for the distribution of orientation angles is
\begin{equation}
    \partial_t p(\hat{\vec  n}) = -\frac{1}{a}\nabla\cdot\left(\nabla\left(\vec w\cdot \hat{\vec n}\right)p\right)+\frac{T}{\gamma}\nabla^2p
    \label{eq:fokkerplanck}
\end{equation}
\subsection{Linear response}
\noindent Linear dynamics of $\vec m$ corresponds to the situation where the distribution $p(\hat{\vec n})$ deviates weakly from isotropic, which is equivalent to assuming $\frac{|\vec w|\gamma}{aT}\ll 1$. The equation of motion for $\vec m$ can be found by multiplying equation (\ref{eq:fokkerplanck}) by $\vec{\hat n}$ and integrating over the unit sphere:
\begin{equation}
    \partial_t \vec m = -2\frac{T}{\gamma}\vec m - \frac{1}{a}\int\nabla\cdot \left(\nabla\left(\vec w\cdot \vec{\hat n}\right)p\right)\vec{\hat n} d\Omega\;.
\end{equation}\\
\noindent Here, we used $\nabla^2\vec{\hat n} = -2\vec{\hat n}$ to simplify the Laplacian term. The integral on the right-hand side can be performed by parts in spherical coordinates, and then using $\langle P_2 \rangle = \frac{1}{2}\int_0^{\pi}P(3\cos(\theta)^2-1)d\Omega$ we get
\begin{equation}
    \partial_t \vec m = -2\frac{T}{\gamma}\vec m + \frac{2\vec w}{3}\left(1-\langle P_2\rangle\right)
\end{equation}
In the linear response regime, the distribution in the second term should be assumed to be isotropic, leading to  $\langle P_2 \rangle = 0$ and we get
\begin{equation}
    \partial_t \vec m = -2\frac{T}{\gamma}\vec m + \frac{2\vec w}{3a}
    \label{eq:linear_eqmot_mag_si}
\end{equation}
which is equation (\ref{eq:eqmot_mag_linear}) used in the main text.
\subsection{Beyond linear response}
Going beyond the linear response regime, we still assume the distribution to be axially symmetric about $\vec w$. In this approximation, the Fokker-Planck equation (\ref{eq:fokkerplanck}) can be evaluated more exactly by considering the time-dependence of each Legendre mode $m_l = \langle P_l(\cos(\theta))\rangle$, where $\cos\theta = \hat{\vec w}\cdot \hat{\vec n}$. To do this, we multiply equation (\ref{eq:fokkerplanck}) by $P_l$ and integrate over the unit sphere. Because of axial symmetry, only the $\hat{\vec z}$ component is relevant:
\begin{equation}
    \tau\partial_t m_l = -l(l+1) m_l - \frac{|\vec w|\tau}{a}\int\nabla\cdot \left(\nabla\left(\cos(\theta)\right)p\right)P_l d\Omega\;,
\end{equation}
where we have inserted the definition $\tau = \gamma/T$. The integral on the right-hand side can be evaluated by integration by parts, which leads to the following sequence of differential equations for $m_l$:
\begin{equation}
    \frac{\tau}{l(l+1)}\partial_t m_l = -m_l + \frac{|\vec w|\tau/a}{2l+1}\left(m_{l-1}-m_{l+1}\right)
\end{equation}
This should be complemented with the initial condition that $m_0 = 1$, ensuring that this can be solved sequentially. Note that the equation for $m_1$ involves $m_2$, a fact which is negligible in the linear response treatment, but important beyond linear response. 

In equilibrium, the time derivatives are all $0$, and the resulting recurrence relation for the equilibrium $m_l^{eq}$ can be identified as the Bessel recurrence relation. Requiring regularity at $w = 0$, one can obtain the following result:
\begin{equation}
    m_l^{eq} = \sqrt{\frac{\pi \alpha}{2}}\frac{I_{l+1/2}(\alpha)}{\sinh{\alpha}}\;,
\end{equation}
where $\alpha = |\vec w|\tau/a$, and $I_{\nu}$ is the modified Bessel function of the first kind. In particular, if $l=1$, we obtain the well-known result $m_1^{eq} = \coth{\alpha}-1/\alpha = \mathcal L(\alpha)$, also known as the Langevin function. 

The dynamical behavior of this system can be investigated perturbatively in the small parameter $\alpha$. Indeed, the leading behavior for the moment $m_l$ is proportional to $\alpha^l$. Therefore, we may consider the leading dynamical behavior by truncating the sequence after $l=2$:
\begin{equation}
\begin{split}
    \tau\partial_t m_1 &= -2m_1 + \frac{2}{3}\alpha(1-m_2)\\
    \tau\partial_t m_2 &= -6m_2 + \frac{6}{5}\alpha m_1\;.
\end{split}
\end{equation}
These can be combined into one equation for $\vec m = m_1\hat{\vec z}$:
\begin{equation}
    \frac{3}{2}\tau^2\partial_t^2 \vec{m}- 12\tau\partial_t \vec{m}- \left(18 + \frac{6}{5}\alpha^2\right)\vec m = -6\alpha\hat{\vec z}
    \label{eq:2ndorder_meq}
\end{equation}
This is the equation for an overdamped harmonic oscillator, around the equilibrium value $\vec m_{eq} = \frac{\alpha}{3 + \alpha^2 /5}\hat{\vec z}\simeq \left(\frac{\alpha}{3} - \frac{\alpha^3}{45}\right)\hat{\vec z}$. The oscillator is overdamped for any $\alpha<5$, and since we are in the regime $\alpha \ll 1$ we can safely neglect inertia.

In the main text, we simplify this overdamped equation of motion by writing it as 
\begin{equation}
    \tau\partial_t \vec{m} = 2\left(\vec m_{eq}(\vec w)-\vec m\right)\;,
    \label{eq:nonlin_mag_eq_si}
\end{equation}
where $\vec m_{eq} = \frac{\vec w \tau}{3 a}\left(1 - \frac{\left(\vec w\tau/a\right)^2}{15}\right)$
\section{Derivation of equations of motion in terms of $\vec u$ and $\vec w$}
\label{sec:uw_eqmot_derivation}
In the main text, we define the velocity fields $\vec w = \vec v^{\mathrm{p}} - \vec v^{\mathrm{s}}$ and $\vec u_{\omega} = \left. \left(\eta^{\mathrm{p}}_{\omega}\vec v^{\mathrm{p}}_{\omega}+\eta^{\mathrm{s}}\vec v^{\mathrm{s}}_{\omega}\right)  \right/ \left(\eta^{\mathrm{p}}_{\omega}+\eta^{\mathrm{s}}\right)$. Here we show how to transform the two-fluid equations of motion into these new variables. We begin with the force-balance equations
\begin{subequations}
\label{eq:eqmot_forcebalance_fluids}
\begin{align}
\zeta(\vec v^{\mathrm{p}}_{\omega} - \vec v^{\mathrm{s}}_{\omega}) &= \eta^{\mathrm{p}}_{\omega}\nabla^2\vec v^{\mathrm{p}}_{\omega} -K \nabla \delta\phi_{\omega} - \phi_0\nabla P_{\omega} + \vec F^{\mathrm{p}}_{\omega}
\label{eqmot_vp_si}\\
\zeta(\vec v^{\mathrm{s}}_{\omega} - \vec v^{\mathrm{p}}_{\omega}) &= \eta^{\mathrm{s}}\nabla^2\vec v^{\mathrm{s}}_{\omega} - (1-\phi_0)\nabla P_{\omega} + \vec F^{\mathrm{s}}_{\omega}\;.
\label{eqmot_vs_si}
\end{align}
\end{subequations}
To get the force-balance equation for $\vec w$, we divide equation (\ref{eqmot_vp_si}) by $\eta^{\mathrm{p}}$ and equation (\ref{eqmot_vs_si}) by $\eta^{\mathrm{s}}$ before taking their difference. This gives us
\begin{equation}
    \begin{split}
   & \zeta\left(\frac{1}{\eta^{\mathrm{p}}_{\omega}}+\frac{1}{\eta^{\mathrm{s}}}\right) \vec w_{\omega}   =  \nabla^2\vec w_{\omega} + \left( \frac{1}{\eta^{\mathrm{p}}_{\omega}} + \frac{1}{\eta^{\mathrm{s}}} \right) f \rho \vec m_{\omega} \\
    & \ \ \ \ \ \ \ \ \ \ \ +  \left(\frac{1-\phi_0}{\eta^{\mathrm{s}}} - \frac{\phi_0}{\eta^{\mathrm{p}}_{\omega}} \right) \nabla P_{\omega} - \frac{K}{\eta^{\mathrm{p}}_{\omega}}\nabla\delta\phi_{\omega} \;.
    \label{eq:eqmot_w_full}
\end{split}
\end{equation}
The equation for $\vec u$ is obtained by simply taking the sum of equations (\ref{eqmot_vp_si}) and (\ref{eqmot_vs_si}), and using the fact that in our model $\vec F^{\mathrm{p}} + \vec F^{\mathrm{s}} = 0$:
\begin{equation}
    \left( \eta^{\mathrm{p}}_{\omega} + \eta^{\mathrm{s}} \right) \nabla^2\vec u_{\omega}  =    K \nabla\delta \phi_{\omega} + \nabla P_{\omega} \;.
    \label{eq:eqmot_u_full}
\end{equation}
Next we turn our attention to the equation of continuity
\begin{equation}
\label{eq:eqmot_cont_fluids}
i\omega \delta\phi_{\omega} = \phi_0\nabla\cdot \vec v^{\mathrm{p}}_{\omega} = - (1-\phi_0) \nabla\cdot \vec v^{\mathrm{s}}_{\omega}\;.
\end{equation}
In the above, we solve for $\nabla\cdot \vec v^{\mathrm{s}},\ \nabla\cdot \vec v^{\mathrm{p}}$ and insert them into the definitions $\nabla\cdot\vec w = \nabla\cdot \vec v^{\mathrm{p}} - \nabla\cdot \vec v^{\mathrm{s}}$, and $\nabla\cdot\vec u = \left(\eta^{\mathrm{p}}\nabla\cdot \vec v^{\mathrm{p}} + \eta^{\mathrm{s}}\nabla\cdot \vec v^{\mathrm{s}}\right)/(\eta^{\mathrm{p}} + \eta^{\mathrm{s}})$. This yields the continuity equation in terms of the two new fields
\begin{subequations}
\begin{align}
    \nabla \cdot \vec w_{\omega} &= i\omega\delta\phi_{\omega}\left(\frac{1}{\phi_0} + \frac{1}{1-\phi_0}\right) \ ,
    \label{eq:cont_w}\\
    \left( \eta^{\mathrm{p}}_{\omega} + \eta^{\mathrm{s}} \right)  \nabla \cdot \vec u_{\omega} &= i\omega\delta\phi_{\omega}\left(\frac{\eta^{\mathrm{p}}_{\omega}}{\phi_0} - \frac{\eta^{\mathrm{s}}}{1-\phi_0}\right) \ .
    \label{eq:cont_u}
\end{align}
\end{subequations}
We now have the full set of equations of motion, in terms of the new velocity fields $\vec u,\vec w$.

\section{Full derivation of equation (\ref{eq:eqmot_longitudinal}, main text) for longitudinal case, as well as more general equation (\ref{eq:eqmot_nonlinear}, main text)}
\label{sec:full_derivation}
In the main text, we used equations (\ref{eq:transverse_flows},\ref{eq:eqmot_longitudinal}, main text) for the transverse and longitudinal flows in the linear regime. Later in the paper, when we turned our attention to non-linear dynamics, we used equation (\ref{eq:eqmot_nonlinear}, main text). Here we will derive a more general result from which both of the linear-response equations of motion, as well as the nonlinear equation of motion may be derived as a particular case, after inserting the appropriate dynamics for $\vec m$:
\begin{equation}
\begin{split}
    -i\omega\tau&\left(1 + \lambda^2\nabla\times\nabla\times - \lambda_s^2\nabla\nabla\cdot\right)\vec w_{\omega} \\
    &-2\lambda_d^2\nabla\nabla\cdot \vec w_{\omega} = 
    -i\omega \frac{f\rho\tau}{\zeta}\vec m_{\omega}\;.
    \label{eq:general_eqmot_w}
\end{split}
\end{equation}
This is the linear response relation describing the response of $\vec w$ to the forcing $\vec m$. 

We begin the derivation of formula (\ref{eq:general_eqmot_w}) with the force-balance equation of motion derived above
\begin{equation}
\begin{split}
    &\left(\zeta\left(\frac{1}{\eta^{\mathrm{p}}_{\omega}} + \frac{1}{\eta^{\mathrm{s}}}\right) - \nabla^2 \right)\vec w_{\omega} = \left(\frac{1}{\eta^{\mathrm{p}}} + \frac{1}{\eta^{\mathrm{s}}}\right)f\rho \vec m_{\omega}\\
    &-\left(\frac{\phi_0}{\eta^{\mathrm{p}}_{\omega}} - \frac{1-\phi_0}{\eta^{\mathrm{s}}}\right)\nabla P_{\omega} - \frac{K}{\eta^{\mathrm{p}}_{\omega}}\nabla\delta\phi_{\omega}\;.
\end{split}
\end{equation}

We then eliminate the pressure gradient $\nabla P_{\omega}$ by solving for it in equation (\ref{eq:eqmot_u_full}). Notice that the divergence-free part of $\vec u$ does not couple to any other fields, so we can safely assume $\nabla\times \vec u = 0$. Then, using the identity
\begin{equation}
    \nabla^2\vec v = \nabla\left(\nabla\cdot \vec v\right) - \nabla\times\left(\nabla\times \vec v\right)\;,
    \label{eq:vector_identity_si}
\end{equation}
which is valid for any vector field $\vec v$, we can write $\nabla^2\vec u_{\omega} = \nabla\left(\nabla\cdot \vec u_{\omega}\right)$. Thus we obtain
\begin{equation}
\begin{split}
    &\nabla P_{\omega} = -K\nabla \delta\phi_{\omega} + \left(\eta^{\mathrm{p}}_{\omega} + \eta^{\mathrm{s}}\right)\nabla^2 \vec u_{\omega}\\
    & = \left(-K + i\omega\left(\frac{\eta^{\mathrm{p}}_{\omega}}{\phi_0} - \frac{\eta^{\mathrm{s}}}{1-\phi_0}\right)\right)\nabla \delta\phi_{\omega}
\end{split}
\end{equation}
where we used the continuity equation (\ref{eq:cont_u}). The equations can be closed by relating $\delta \phi$ back to $\vec w$ using equation (\ref{eq:cont_w}). We have $\nabla\delta\phi_{\omega} = \frac{\phi_0(1-\phi_0)}{i\omega}\nabla\left(\nabla\cdot\vec w_{\omega}\right)$, thus leading to the full equation of motion
\begin{equation}
\begin{split}
    &-i\omega \left(\zeta-\frac{\eta^{\mathrm{p}}\eta^{\mathrm{s}}}{\eta^{\mathrm{p}}+\eta^{\mathrm{s}}}\nabla^2\right)\vec w_{\omega} = -i\omega f\rho\vec m_{\omega}\\
    &+ K\phi_0(1-\phi_0)^2\nabla\left(\nabla\cdot\vec w_{\omega}\right) \\
    &- i\omega\frac{\left((1-\phi_0)\eta^{\mathrm{p}} - \phi_0\eta^{\mathrm{s}}\right)^2}{\eta^{\mathrm{s}} + \eta^{\mathrm{p}}}\nabla\left(\nabla\cdot \vec w_{\omega}\right)\;.
\end{split}
\end{equation}
We can collect terms and rewrite the equation as follows
\begin{equation}
\begin{split}
    -i\omega\tau&\left(1 - \lambda^2\nabla^2 - (\lambda_s^2-\lambda^2)\nabla\nabla\cdot\right)\vec w_{\omega} = \\
    &2\lambda_d^2\nabla\nabla\cdot \vec w_{\omega} -i\omega \frac{f\rho\tau}{\zeta}\vec m_{\omega}\;,
\end{split}
\label{eq:geneal_eqmot_nocross}
\end{equation}
where we have defined the length scales
\begin{equation} 
\begin{split} 
\lambda^2 &= \frac{\eta_{\omega}^{\mathrm{p}} \eta^{\mathrm{s}} /\zeta}{\eta_{\omega}^{\mathrm{p}} + \eta^{\mathrm{s}}} \simeq \frac{\eta^{\mathrm{s}}}{\zeta}\;,\\
\lambda_s^2 & = \frac{\eta_{\omega}^{\mathrm{p}} \eta^{\mathrm{s}} + \left( \eta_{\omega}^{\mathrm{p}} (1- \phi_0) - \eta^{\mathrm{s}} \phi_0 \right)^2}{\zeta (\eta_{\omega}^{\mathrm{p}} + \eta^{\mathrm{s}})} \simeq \\ & \simeq \frac{  \eta_{\omega}^{\mathrm{p}} \left(1- \phi_0 \right)^2 }{\zeta }\;, \\ & \text{and} \\ \lambda_d^2 & = \frac{K \phi_0 \left( 1 - \phi_0 \right)^2\gamma}{ 2\zeta T }\;.
\end{split}
\label{eq:lengthscales}
\end{equation}
Using the identity (\ref{eq:vector_identity_si}), equation (\ref{eq:geneal_eqmot_nocross}) can be finally transformed into the desired equation (\ref{eq:general_eqmot_w}). It is the general equation of motion for the velocity field $\vec w$, agnostic to the specific dynamics that $\vec m$ obeys. It is valid for both longitudinal and transverse modes, as well as any combination of them.
\subsection{Derivation of the linearized equation of motion}
\label{sec:linear_derivation}

As we mentioned, equation  (\ref{eq:general_eqmot_w}) represents the linear response of $\vec w$ given some source $\vec m$. It can be formally solved for the Green's function of the velocity field for a given orientation field. The physical description of the system is complete once we introduce the feedback of $\vec w$ on $\vec m$, which is done through the linear equation (\ref{eq:linear_eqmot_mag_si}), as long as we are in the linear response regime. Altogether, this gives a closed linear equation of motion for $\vec w$:
\begin{equation}
\begin{split}
    &\left(i \omega \tau \right)^2 (1+\lambda^2\nabla\times\nabla\times-\lambda_s^2\nabla\nabla\cdot) \vec w_{\omega}  \\ & - 2 \left(i \omega \tau \right) \left(  1 - \frac{f \rho \gamma}{3 a \zeta T} +\lambda^2\nabla\times\nabla\times-  (\lambda_s^2+\lambda_d^2)\nabla\nabla\cdot \right) \vec w_{\omega} \\
    &- 4 \lambda_d^2 \nabla\nabla\cdot\vec w_{\omega} = 0 \ .
\end{split}
\label{eq:eqmot_linear_si}
\end{equation}
The beauty of this equation is that it automatically produces equations (\ref{eq:transverse_flows}, main text) and (\ref{eq:eqmot_longitudinal}, main text). For the transverse case, when we take $\vec w = \vec w_{\perp}$, since $\nabla\cdot \vec w_{\perp} = 0,\; \nabla\times\nabla\times \vec w_{\perp} = -\nabla^2\vec w_{\perp}$, this produces (\ref{eq:transverse_flows}, main text). Conversely, when we take the longitudinal component $\vec w = \vec w_{\parallel}$, then  $\nabla\nabla\cdot \vec w_{\parallel} = \nabla^2\vec w_{\parallel}$, and $\nabla\times\vec w_{\parallel} = 0$, and we get (\ref{eq:eqmot_longitudinal}, main text). 

It is worth noting that this structure, of one equation describing response and the other feedback, each with their own timescale, is strongly reminiscent of the structure of the Lotka-Volterra equations for predator-prey dynamics. As we have noted in the main text, the resonant timescale of the oscillator we obtain for the longitudinal modes is the geometric mean of the two underlying relaxation times, just as in the Lotka-Volterra model \cite{lotka1910contribution,volterra1928variations}.

Instead of excluding $\vec m$ and writing the equation of motion for $\vec w$ (\ref{eq:eqmot_linear_si}), we can equally well exclude $\vec w$ and write the equation of motion for $\vec m$. This happens to have identically the same form as equation (\ref{eq:eqmot_linear_si}). Mathematically, this is due to the fact that the differential operators that relate $\vec m$ and $\vec w$ commute with one another.

We can also derive a closed equation of motion for $\delta\phi$. Again, we begin with the linear response relation for $\delta \phi$ given $\vec m$, which may be derived by taking the divergence of equation (\ref{eq:general_eqmot_w_si}) and using the continuity equation (\ref{eq:cont_w})
\begin{equation}
\begin{split}
    \left(1-\lambda_s^2\nabla^2\right)\tau\partial_t\delta\phi = &2\lambda_d^2\nabla^2\delta\phi\\
    &+ \frac{f\rho\tau\phi_0(1-\phi_0)}{\zeta}\nabla\cdot \vec m\;.
\end{split}
\label{eq:eqmot_phi}
\end{equation}
Within linear response, taking the feedback equation in linear form (\ref{eq:linear_eqmot_mag_si}), this produces
\begin{equation}
\begin{split}
    (1&-\lambda_s^2\nabla^2)\tau^2\partial_t^2\delta\phi-4\lambda_d^2\nabla^2\delta\phi \\
    &+2\left(1-\frac{f\rho\tau}{3a\zeta} - (\lambda_s^2 + \lambda_d^2)\nabla^2\right)\tau\partial_t\delta\phi = 0\;.
\end{split}
\end{equation}
As before, at any wave vector $q$, this is an oscillator equation with friction term affected by the force.  Analysis of this equation, therefore, leads to the same conclusions as before.

\subsection{Nonlinear regime}
\label{sec:nonlinear_derivation}
Beyond linear response when nonlinearities are at play, we cannot resort to Fourier modes, so we must work with a version of equation (\ref{eq:general_eqmot_w}) in the time domain
\begin{equation}
\begin{split}
    \left(1 + \lambda^2\nabla\times\nabla\times - \lambda_s^2\nabla\nabla\cdot\right)\tau\partial_t\vec w = &2\lambda_d^2\nabla\nabla\cdot \vec w \\
    &+\frac{f\rho\tau}{\zeta}\partial_t\vec m\;.
\end{split}
\label{eq:general_eqmot_w_si}
\end{equation}
We formally write the solution of the nonlinear equation for $\vec m$ as 
\begin{equation}
    \vec m = \left( 2 + \tau \partial_{t} \right)^{-1} 2 \vec m_{\mathrm{eq}}(\vec w)\;,
\end{equation}
plug this solution into equation (\ref{eq:general_eqmot_w_si}), and then use the fact that the operator $\left( 2 + \tau \partial_{t} \right)$ commutes with both spatial and time derivatives in equation (\ref{eq:general_eqmot_w_si}).  As a result, we arrive at 
\begin{equation}
\begin{split}
 & \tau^2 \partial_t^2\left[ 1 - \lambda_s^2 \nabla \nabla \cdot + \lambda^2 \nabla \times \nabla \times \right] \vec w - 4 \left[ \lambda_d^2 \nabla \nabla \cdot \right] \vec w   \\ & + 2 \tau\partial_t \left[ 1  - \left( \lambda_s^2+\lambda_d^2 \right)\nabla \nabla \cdot + \lambda^2 \nabla \times \nabla \times  - \right. \\ & \left. \ \ \ \ \ \ \ \ \ \ \ \ \ \ \ \ \ \ \ \ \ - \frac{ f \rho \gamma}{3 a \zeta T} \left( 1 - \frac{ \tau^2}{15 a^2} \vec w^2 \right) \right] \vec w = 0 \ ,
\end{split}
\label{eq:nonlin_eqmot_si}
\end{equation}
which is equation (\ref{eq:eqmot_nonlinear}) in the main text.

\subsection{List of possible nonlinearities}
\label{sec:nonlin_list}
In our analysis of the nonlinear regime above, we investigated the effects of the saturation of the orientation field $\vec m$, which cannot take values $|\vec m| > 1$. We deem this to be an important nonlinearity to consider, as otherwise the system quickly diverges into states which violate the very definition of $\vec m$ as an average of unit vectors, rendering the model inconsistent. In addition however, there are a number of deviations from the linear response regime which could be taken into account, but which we choose to neglect for simplicity. These include:
\begin{itemize}
    \item The advection of force dipoles by the surrounding fluid flow, which would result in a term proportional to $\vec v^{\mathrm{p}} \cdot \nabla\vec m$ added to Equation (\ref{eq:nonlin_mag_eq_si}). 
    \item The nonlinear osmotic pressure $\Pi$ due to large variations in concentration, proportional to $\delta\phi^2$ and higher powers.
    \item Nonlinear rheology (dependence of stress tensor $\sigma^{\mathrm{p}}$ on velocities), such as shear-thinning or thickening effects, as well as non-local rheological response (so-called $q-$dependent rheology [56]). 
    \item Nematic contribution to the stress tensor, proportional to $\vec w\vec w$.
    \item Active nematic contribution to the stress tensor, proportional to $f\vec w \vec w$. 
    \item Dependence of activity on density, as has been observed in the case of bacterial swarms for example. This would have a generic nonlinear effect on the microscopic forcing of the dipoles $f(\delta\phi)$.
    \item Extra osmotic pressure due to activity, be it due to resulting ATP concentration gradients or other chemical fuels and waste resulting from activity.
\end{itemize}
We choose to neglect these so that our model may be tractable analytically, however they may be included in future numerical studies of this model. 

\subsection{Stability and  conservation of mass}
\label{sec:conservation_phi}

Although by construction our equations describe only the redistribution of chromatin driven by motors, and do not involve either change in the amount of material or spontaneous motion of chromatin, it is technically useful and important to see how these properties are implemented in the final equations of motion, like Eq. (\ref{eq:nonlin_eqmot_si}). Furthermore, it will be useful for us to ensure that our numerical scheme detailed in Section \ref{sec:numerics} indeed satisfies these constraints.

Consider, for instance, the linear response equation for $\delta \phi$ (\ref{eq:eqmot_phi}). If there is no drive, i.e. $\vec m = 0$, but $\delta\phi$ happens to be nonzero at $t=0$, then (\ref{eq:eqmot_phi}) guarantees that $\delta\phi$ will decay stably to $0$. This follows from the fact that the Laplacian operator has negative eigenvalues. For instance, in an infinite domain where we can write $\nabla^2\rightarrow -q^2$, we would have
\begin{equation}
    \delta\phi_q(t) = \delta\phi_q(0)\exp\left(-\frac{2\lambda_d^2 q^2}{1+\lambda_s^2q^2}t\right)\;.
\end{equation}

Consider now the more interesting case where there is a drive, $\vec m \neq 0$. Suppose first that the domain is very large but $\vec m$ is only located in some part of this domain, while far away both $\vec m$ and $\delta \phi$ are $0$. Then, integrating equation (\ref{eq:eqmot_phi}) over the whole volume gives 
\begin{equation}
    \partial_t\int \delta\phi dV = 0\;,
\end{equation}
which means the total amount of polymer material is conserved, as expected.

In the case of a finite domain $\Omega$ where activity may happen close to the boundary, we still expect a boundary condition $\vec w = 0,\; \vec m = 0$ at the boundary (or, if there is hydrodynamic slip on the boundary, then only the normal components are $0$, which does not affect our conclusions). Then, integrating equation (\ref{eq:eqmot_phi}) over $\Omega$, we are left with
\begin{equation}
\begin{split}
    \tau&\partial_t \int_{\Omega}\delta\phi \text{d}V = \oint_{\partial \Omega} \left(\lambda_d^2 + \lambda_s^2\tau \partial_t\right)\nabla\delta\phi\cdot \text{d}\vec S\\
    &= \frac{\phi_0(1-\phi_0)^2\tau}{\zeta}\oint_{\partial \Omega}\left(K\nabla\delta\phi - \eta^{\mathrm{p}}\nabla\nabla\cdot\vec v^{\mathrm{p}}\right)\cdot \text{d}\vec S\;,
\end{split}
\label{eq:cons_phi}
\end{equation}
where we have used divergence theorem on the right-hand side integral, followed by using the continuity equation (\ref{eq:cont_w}). The term on the right-hand side must therefore be $0$ to guarantee the conservation of $\delta\phi$. This is seen by remembering the force-balance condition for the polymer at the boundary: since $\vec v^{\mathrm{p}} = \vec m = 0$ at the boundary, the only forces are due to viscosity and osmotic pressure, which must exactly cancel out. Thus, the integrand in the right-hand-side of (\ref{eq:cons_phi}) is exactly $0$ everywhere at the boundary.

Finally, the numerical scheme we show in Section \ref{sec:numerics} has no boundaries and assumes a periodic domain, so it will automatically conserve $\int\delta\phi dV$.

\section{Numerical Solutions}
\label{sec:numerics}

To investigate the solutions to the nonlinear equation of motion (\ref{eq:nonlin_eqmot_si}), we wrote a simple numerical scheme to solve the system in one dimension. We write the equations using a non-dimensional version of the velocity field $\tilde w(x,t) = \frac{\tau}{3a} w(x,t)$. We use the two equations (\ref{eq:nonlin_mag_eq_si}) and (\ref{eq:general_eqmot_w_si}) instead of combining them, which allows us to numerically integrate only first-order differential equations in time. In one-dimensional form, these equations read
\begin{equation}
\begin{split}
    (1-\lambda_s^2\partial_x^2)\partial_t \tilde w &= 2\lambda_d^2\partial_x^2 \tilde w + (\epsilon+1)\partial_t m\\
    \partial_t m &= -2m + 2\tilde w\left(1-\beta\frac{3}{5}\tilde w^2\right)\;.
\end{split}
\label{eq:numerical_eqs}
\end{equation}
Here, $\beta$ is a parameter which we set to $0$ or $1$ depending on whether we want to consider nonlinear effects. We have also set the characteristic time $\tau = 1$. We solve these equations using an explicit forward time-stepping scheme, and treat the spatial derivatives with Fast Fourier Transform by assuming the domain is periodic. It is worth noting that since these two equations are equivalent to one second-order differential equation in time for $\tilde w$, we must specify two initial conditions. Either we set $\tilde w(x,t=0),m(x,t=0)$, or one of these must be specified along with its time derivative. For all of the solutions below, we set the screening length to be much smaller than the domain size, $\lambda_s/L = 10^{-4}$, since we are interested in the large-scale near-critical dynamics of this system.

\begin{figure}[t]
    \centering
    \includegraphics[width=\linewidth]{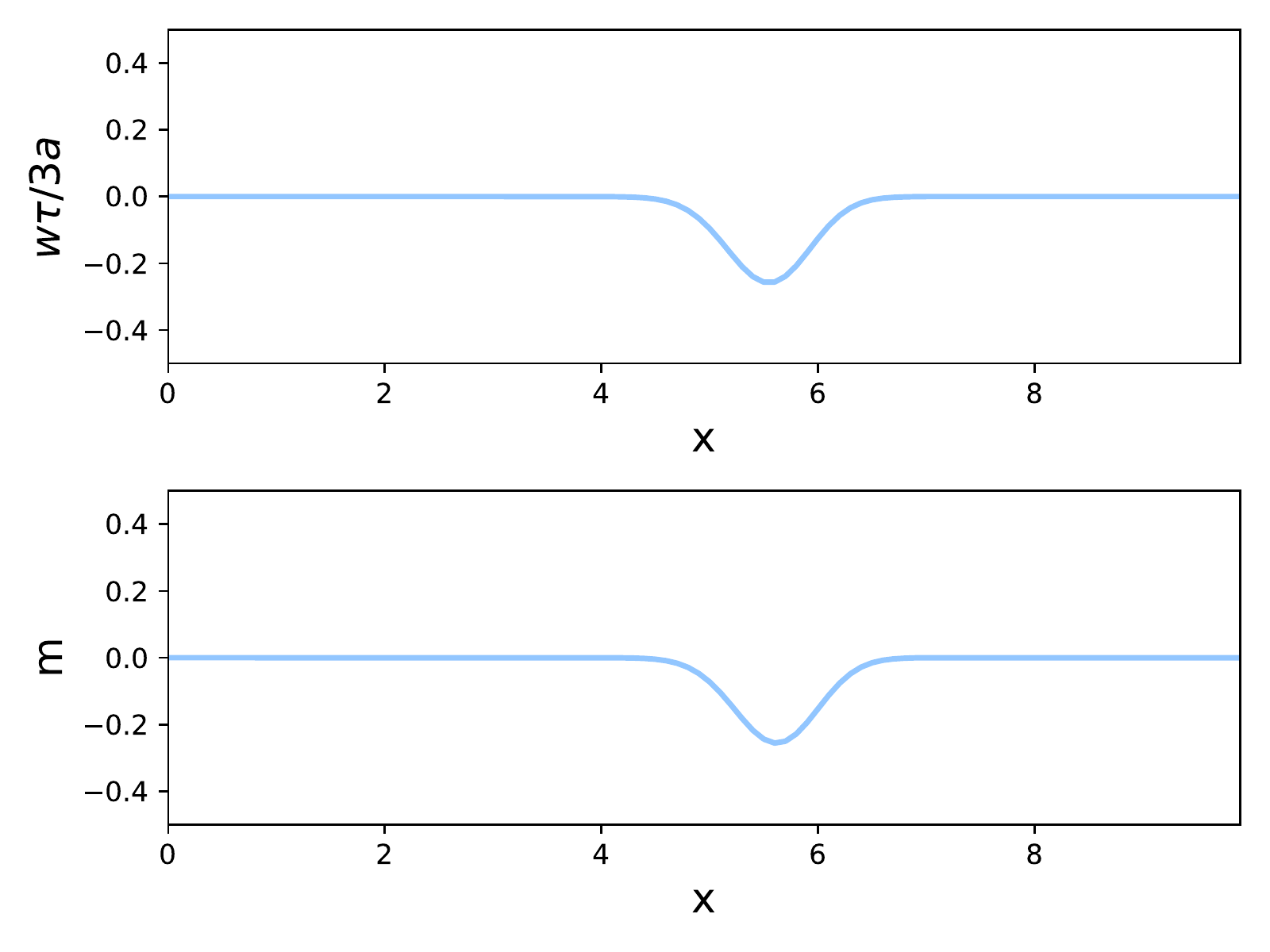}
    \caption{Initial condition  for $m,\tilde w$ used in Supplemental Movie 2, which leads to a conserved wave-packet moving to the left at constant speed.}
    \label{fig:conserved_wavepacket_fig}
\end{figure}
\subsection{Linear dynamics}

First we set $\beta = 0$ and investigate the linear dynamics. As expected, when $\epsilon > 0$ the solutions diverge in time, but keeping this parameter close to $0$ (we chose $\epsilon = 0.02$), we observe interesting transient dynamics. We initialize $\tilde w(x,t=0) = 0$, and set $m(x,t=0)$ to be a localized perturbation in the form of the derivative of a Gaussian with width $0.2$, while the size of the whole domain is $L = 10$. This initial condition does not select a direction of propagation, which is why we observe it splitting into two wave packets which move away from each other at a constant speed. In the main text, we identified this speed as being set by the combination $\lambda_d/\tau$. This is shown in Supplemental Movie 1. 

We also initialized the dynamics with an initial condition which does set the direction of propagation, by initializing $\tilde w$ and $m$ as shown in Fig. \ref{fig:conserved_wavepacket_fig}. When the system is thus initialized, the packet moves to the left at a constant speed and its shape is conserved. We first considered these dynamics for small $\epsilon$ which leads to instabilities developing very slowly. Thus, this wavepacket keeps its shape for the duration of the numerical integration. This is shown in Supplemental Movie 2.

\subsection{Nonlinear dynamics}

\begin{figure}[t]
    \centering
    \includegraphics[width=\linewidth]{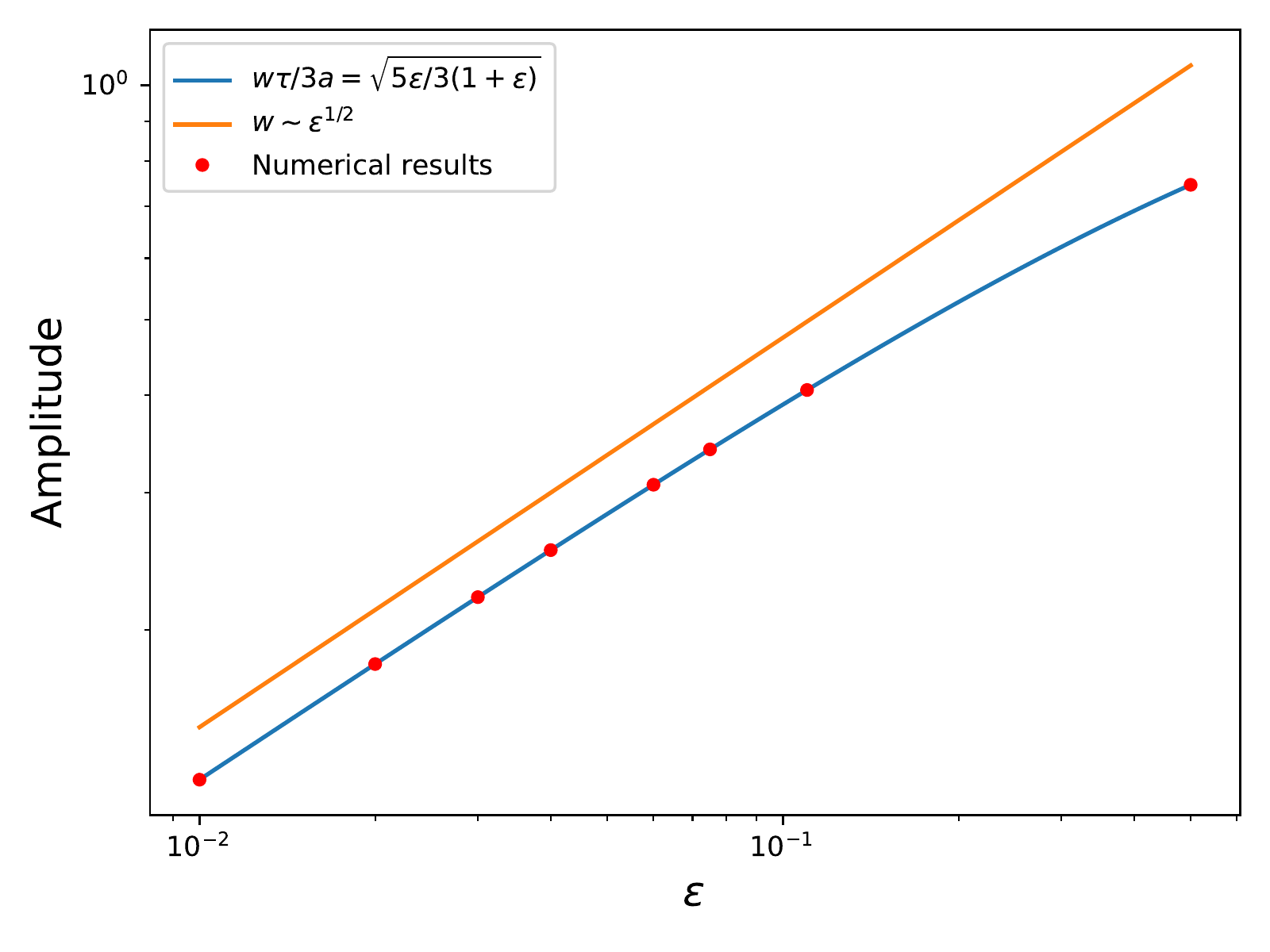}
    \caption{Scaling of the wave amplitude with the critical parameter $\epsilon$, as determined by numerical integration of equations (\ref{eq:numerical_eqs}). For small values of $\epsilon$, the scaling of the amplitude follows the expected $\epsilon^{1/2}$. At larger values there is a deviation. This can be explained by solving for the steady-state of the equations of motion, which give an expected amplitude of $\sqrt{5\epsilon/3(1+\epsilon)}$.}
    \label{fig:amplitude_scaling}
\end{figure}

After turning on the nonlinearity, we increased the critical parameter to $\epsilon = 0.2$ so the system quickly reaches the nonlinear regime. We initialize the fields with $\tilde w = 0$, and $m(x,t=0)$ also corresponding to the derivative of a Gaussian with width $0.2$. After some complex developments, the system settles into a steady evolution where a near-square wave propagates at constant speed, which we checked to be close to $2\lambda_d/\tau$, shown in Fig. \ref{fig:velocity_scaling}. A movie showing the development of such nonlinear waves is shown in Supplemental Movie 3. The amplitude of the waves scales as $\sqrt{\epsilon}$, as shown in Fig. \ref{fig:amplitude_scaling}, where we scanned multiple values of $\epsilon$ and measured the amplitude of the resulting steady waves. When $\epsilon$ gets large, the amplitude slightly deviates from the simple power-law behavior, and instead follows $\tilde w = \sqrt{\frac{5\epsilon}{3(1+\epsilon)}}$. The latter relation can be found by solving for a constant steady-state in the equations (\ref{eq:numerical_eqs}). In Supplemental Movie 3, these predicted amplitudes are shown as black dashed lines.
We also numerically verified that the wave speed in these nonlinear waves scales linearly with $\lambda_d$. Over a range of values for this parameter, we tracked the maximum of a traveling pulse and recorded its velocity. These velocities grow linearly with $\lambda_d$ as expected, shown in Fig. \ref{fig:velocity_scaling}.
\begin{figure}[t]
    \centering
    \includegraphics[width=\linewidth]{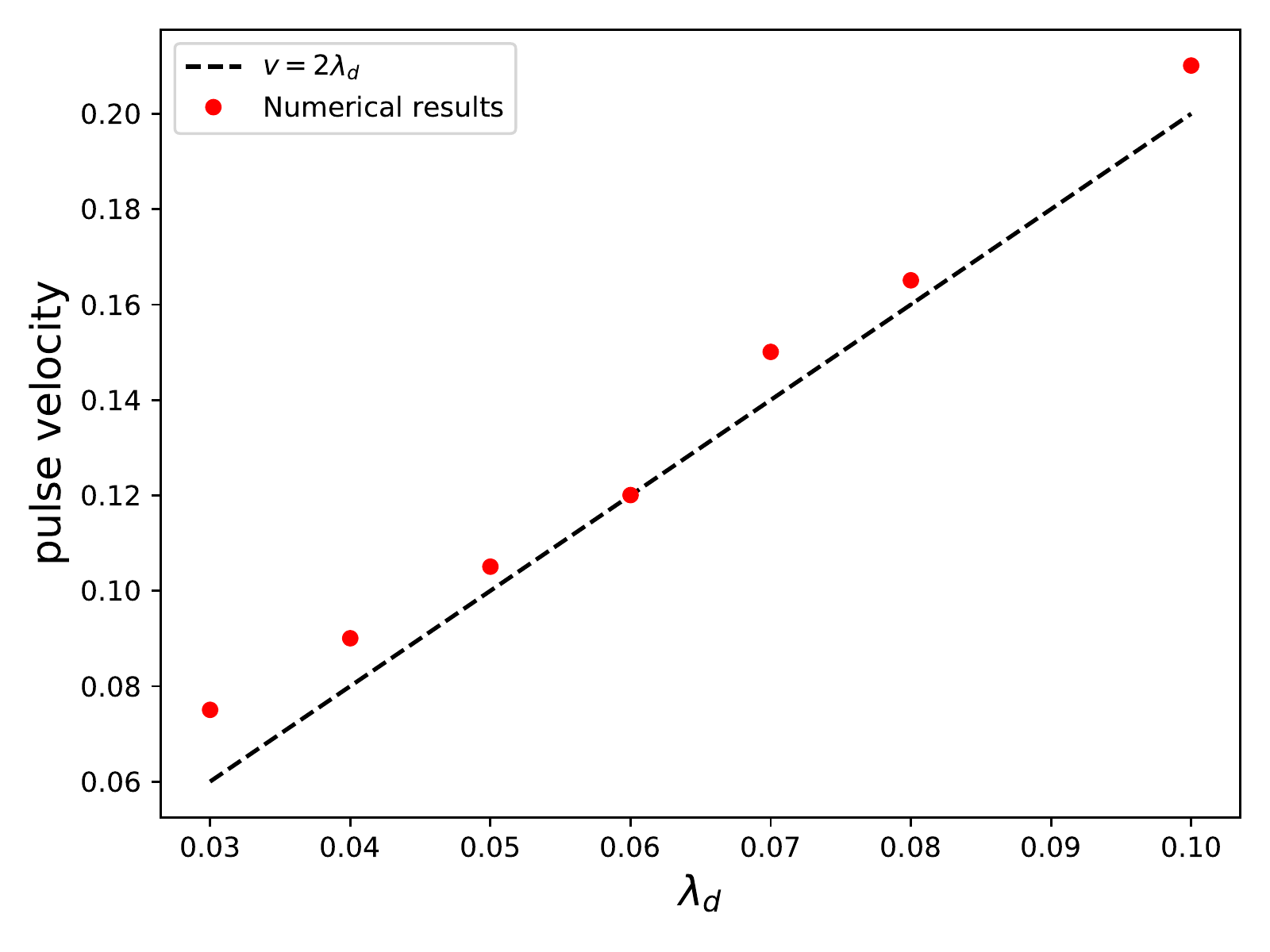}
    \caption{Scaling of pulse velocity with the osmotic lengthscale $\lambda_d$, as measured from numerical integration of (\ref{eq:numerical_eqs}). From our equations of motion we expected the velocity to scale linearly with $\lambda_d$, and indeed here the line $v = 2\lambda_d$ goes through the data, confirming our expectations.}
    \label{fig:velocity_scaling}
\end{figure}
\section{Estimates}
\label{sec:estimates}

In the main text, we make the claim that molecular motors are too far from one another to form long-range order due to their direct contacts. There are approximately $10^5$ RNA polymerase II molecules in a HeLa cell \cite{kimura1999quantitation}, whose nucleus is approximately $\unit[10]{\mu m}$ in diameter, leading to a density around $\unit[10^{2}]{\text{molecules}/\mu m^{3}}$, which corresponds to the average distance of the order of $\unit[\sim 200]{nm}$. In other words, if we take the size of RNA polymerase to be on the order of $\unit[10]{nm}$ in each dimension \cite{milo2015cell}, then we obtain a small volume fraction $\phi_{\mathrm{RNAPII}} \sim 10^{-4}$. This sparseness is also seen, despite some local functional clustering, in superresolution microscopy experiments \cite{zhao2014spatial}.

We had previously estimated \cite{eshghi2022symmetry} the relevant length scales, and we will repeat these estimates here so that this paper may be self-contained. We expect the mesh size, $\lambda$, to range from around $\unit[30]{nm}$ to $\unit[100]{nm}$ \cite{gorisch2003diffusion,solovei2002spatial}. In contrast, the ratio of viscosities $\eta^{\mathrm{p}}/\eta^{\mathrm{s}}$ is harder to estimate. Bare nucleoplasm has been measured to have a viscosity on the same order as that of water \cite{liang2009noninvasive,erdel2015viscoelastic}, $\eta^{\mathrm{s}} \sim \unit[10^{-3}]{Pa\cdot s}$ , whereas a wide range of chromatin viscosities has been measured, $\eta^{\mathrm{p}} \approx \unit[0.6 - 3000]{Pa\cdot s}$  \cite{tseng2004micro,de2007direct,liang2009noninvasive,celedon2011magnetic,hameed2012dynamics,erdel2015viscoelastic,caragine2018surface,eshghi2021interphase,zidovska2020rich}, reflecting in part the complicated nature of this quantity. Thus, experimental ranges for $\eta^{\mathrm{p}}/\eta^{\mathrm{s}}$ lie between $10^2$ and $10^6$. The screening length scale relevant in this paper is $\lambda_s = \eta^{\mathrm{p}}/\zeta = \lambda\sqrt{\eta^{\mathrm{p}}/\eta^{\mathrm{s}}}$. At the upper limit of the estimates, this length scale becomes much larger than the size of the nucleus, making it irrelevant for our system of interest. The lower limit is $\approx \unit[300]{nm}$, which is more consistent with the length scales relevant in the context of chromatin.

To estimate the length scale $\lambda_d$, we assume $K \simeq \frac{T}{\lambda^3}$ \cite{de1979scaling}, and $\gamma \simeq C\eta^{\mathrm{s}} a^3$, where the constant $C$ is an unknown parameter, resulting from the fact that it is unclear whether the motors experiencing the rotational friction $\gamma$ are able to "feel" the polymer viscosity or whether they are small enough that the only relevant viscous dissipation is that of the solvent. As their size is about $20$ nm, comparable to the mesh size \cite{rhodin2003single}, the constant $C$ can be assumed to be $C \ll \eta^{\mathrm{p}}/\eta^{\mathrm{s}} \sim (10^2-10^6)$. From these assumptions, we obtain $\lambda_d \sim \sqrt{\frac{a^3}{\lambda}}\sqrt{C} \ll \lambda_s$, resulting in roughly $\unit[10-20]{nm}$. Finally, we estimate the dipole relaxation time $\tau \simeq C\eta^{\mathrm{s}}a^3/T \sim C \unit[10^{-6}]{s}$, and so the expected speed for traveling polymer waves is on the order of $\lambda_d/\tau \sim \unit[10^7]{ nm/s}$.

\bibliography{sources}
\end{document}